\documentstyle[epsf]{l-aa}
\begin{document}
\newcommand{\m}{$-$}
\newcommand{\lmcv}{LMC\,4}
\newcommand{\civ}{C\,IV}
\newcommand{\kms}{km\,s$^{-1}$}
\newcommand{\snn}{Sk\,$-$66\,100}
\newcommand{\ses}{Sk\,$-$66\,117}
\newcommand{\sea}{Sk\,$-$66\,118}
\newcommand{\ssn}{Sk\,$-$67\,169}
\newcommand{\sns}{Sk\,$-$67\,206}
\newcommand{\sss}{Sk\,$-$67\,166}
\title{C\,IV absorption from hot gas inside supergiant shell \lmcv\ 
observed with HST and IUE 
\thanks{Based on observations obtained with the NASA/ESA {\it Hubble Space 
Telescope} obtained at the Space Telescope Science Institute, 
which is operated by AURA Inc. under NASA contract NAS 5-26555; 
and with the IUE satellite from VILSPA, Spain, jointly operated by 
the NASA, ESA, and PPARC}
}
\subtitle{}
 
\author{
D.J. Bomans \inst{1,2,}\thanks{Feodor Lynen fellow of the Humboldt Foundation}
\and K.S. de Boer \inst{1} 
\and J. Koornneef\, \inst{3} \and E.K. Grebel \inst{1,4}
} 

\institute{
Sternwarte der Universit\"at Bonn, Auf dem H\"ugel 71, D-53121 Bonn, 
Federal Republic of Germany
\and
University of Illinois, Dept. of Astronomy, 1002 West Green St., 
Urbana, IL 61801, U.S.A.
\and
SRON Laboratory for Space Research, Postbus 800, NL-9700AV Groningen, 
The Netherlands
\and
Space Telescope Science Institute, 3700 San Martin Drive, Baltimore, MD 21218, 
U.S.A. 
}
 
\date{received 21 Nov.\, 1995; accepted 11 Jan.\, 1996}

\thesaurus{09.02.1, 09.19.1, 11.09.4, 11.09.5, 11.13.1, 13.21.3}
 
\offprints {D.J. Bomans, USA address, email: {bomans@astro.uiuc.edu}}
 
\maketitle
 
\markboth{Bomans, de Boer, Koornneef, Grebel: C\,IV gas in supergiant shell 
\lmcv}{}

\begin{abstract}
High resolution ultraviolet spectra with HST-GHRS of two stars in the 
direction of the supergiant 
shell \lmcv\ unambiguously show absorption by substantial quantities of \civ\ 
gas at velocities near the the systemic velocity of the LMC.
In combination with the detection of diffuse X-rays 
from the \lmcv\ by ROSAT and other supporting data, this demonstrates that 
the interior of \lmcv\ is filled with tenuous hot gas.\\
\civ\ interstellar absorption is seen over a large velocity range, 
having at least 2 components at about 280 and 320 \kms. 
The strong component at 280 \kms\ has a width of 
40 \kms\ and a column density in the order of $3~10^{13}$ cm$^{-2}$. 
The width of the absorption is best explained by bulk motions 
of \civ -containing gas clouds inside \lmcv. 
These hot clouds or layers around cold clouds 
have to have a relatively high filling factor inside \lmcv\ to fit the 
observations.
The characteristics of the \civ\ gas component at 320 \kms\ 
are such that they 
trace a blast wave from a recent supernova within the \lmcv\ cavity.\\
Galactic \civ\ absorption is also present, as to be expected for these lines 
of sight. 


\end{abstract}

\section{Introduction}

Supergiant shells are important features in the overall structure of 
a number of starforming galaxies \footnote{Large shell structures have 
now been identified in at least 30 nearby galaxies (see e.g. Bomans 1994).}. 
Typically, they are outlined by numerous diffuse and often filamentary HII 
regions which provide evidence for recent star formation at 
the interface between the shell and the ambient interstellar medium.
The ionized shell encloses a space which contains remarkably little  
warm and/or cold gas (for \lmcv\ see the HI map in Rohlfs et al. 1984, 
and the CO map in Cohen et al. 1988).
In the Large Magellanic Cloud (LMC) several such structures are known 
(Goudis \& Meaburn 1978, Meaburn 1980). Of these, the supergiant 
shell \lmcv\ is the largest with  a diameter of over 1.2 kpc. 

A clear connection between the existence of supergiant shells 
and active starformation has been established 
(see e.g. Deul \& den Hartog 1990, Hunter et al. 1993),
but models explaining the origin of the largest of these structures 
are still lacking.
While the combination of winds from hot, young stars and the blasts from 
supernova (SN) explosions from a single starburst event 
(e.g. Mac Low et al. 1989) appears adequate 
to produce the smaller shells, this model fails 
for the largest structures (e.g. Igumentshchev et al. 1990).
Adding the effects of self-propagating star formation to 
the overall energy input may provide a way to create objects 
like \lmcv\, but has not yet been thoroughly modelled.
However, if the supershells were due to radially propagating star formation, 
the young stellar associations and H\,II regions that define the shell 
would be expected to surround an interior containing older stars. 
Observationally, the observed age structure of various stellar groups in 
\lmcv\ (see Vallenari et al. 1993, Bomans et al. 1994, Bomans et al. 1995, and 
Domg\"orgen et al. 1995 for discussions) does not appear to match this prediction. 
A further complication is that the size of the supershells exceeds the 
scale height of a typical galaxy. Therefore, 
blow-out -- i.e., mass and energy loss into the halo -- is  
critical for the understanding of these structures. 
Hydrodynamical simulations of astrophysical plasmas 
(e.g. Suchkov et al. 1994) are consistent with the observation that 
little neutral or low-ionized gas exists within the supergiant shells, 
but require a hot and highly ionized component 
It is this gas that we have been trying 
to definitively establish trough its absorptions in the ultraviolet.

\begin{table*}
\caption[]{Data on the target stars}
\begin{tabular}[]{lllllll}
\hline
Star & RA & DEC & V & Sp.T. & IUE & HST\\
\hline
\ses & 5 30 36 & \m 66 49 & 12.68 & B3 I      & y & n\\
\sea & 5 30 54 & \m 66 55 & 11.81 & B2 Ia$^a$ & y & y\\
\snn & 5 27 49 & \m 66 58 & 13.26 & O6 Ib$^c$ & y & n\\
\sns & 5 34 48 & \m 67 03 & 12.00 & BN0.5 Iaw$^a$ & y & n\\
\ssn & 5 31 54 & \m 67 04 & 12.18 & B1 Ia$^a$ & y & y\\
\sss & 5 31 50 & \m 67 40 & 12.27 & O4 If+$^b$ & y & n\\
\hline
\end{tabular}

Spectral types are from: 
$^a$Fitzpatrick (1988), $^b$Walborn (1977), $^c$Conti et al. (1986)
\end{table*}

A first indication for the presence of highly ionized gas in the LMC was 
found by Koornneef et al. (1979) and de Boer et al. (1980) 
in high resolution spectra obtained with the 
International Ultraviolet Explorer (IUE) satellite. 
Absorption by \civ\ was present in many directions indicating the 
presence of a halo of coronal gas around the LMC (de Boer \& Savage 1980). 
In particular, the presence of \civ\ absorption in spectra of HD 36402 in the 
general direction of \lmcv\ (de Boer \& Nash 1982) suggested the existence 
of gas of at least 10$^5$K. On the basis of the observations on this line of 
sight, a pressure of $nT~=~2~10^4$ cm$^{-3}$\,K was computed for gas inside 
\lmcv\,. 
However, HD 36402 is located at the edge of \lmcv\ within a distinct structure 
(the giant, 100 pc diameter, shell N\,51D), allowing the possibility that this
 \civ\ absorption had a more local origin. A more recent observation by 
Chu \& Mac Low (1990) shows that the inside of the nebula N\,51D is bright in 
soft X-rays, thus confirming the existence of a very hot and highly ionized 
component.

A more definitive search for \civ\ absorptions in \lmcv\ should therefore 
rely on the observations of stars whose lines of sight intersect the bubble 
closer to its center. 
The number of targets satisfying this criterion is very limited, and 
additional requirements are to be put on the spectral types of the stars. 
A high UV continuum flux is needed to provide a sufficient signal to noise 
ratio, but the stars should not be so hot as 
to produce significant amounts of circumstellar \civ. This limits the 
optimal targets to luminous stars, preferably in the spectral types to O9 
to B2. 
Given these constraints, we selected stars inside \lmcv\, for our 
intended science program and added a few more stars observed with IUE 
for different programs. The stars are listed in Table 1.

\begin{figure}
\def\epsfsize#1#2{1.0\hsize}
\centerline{\epsffile{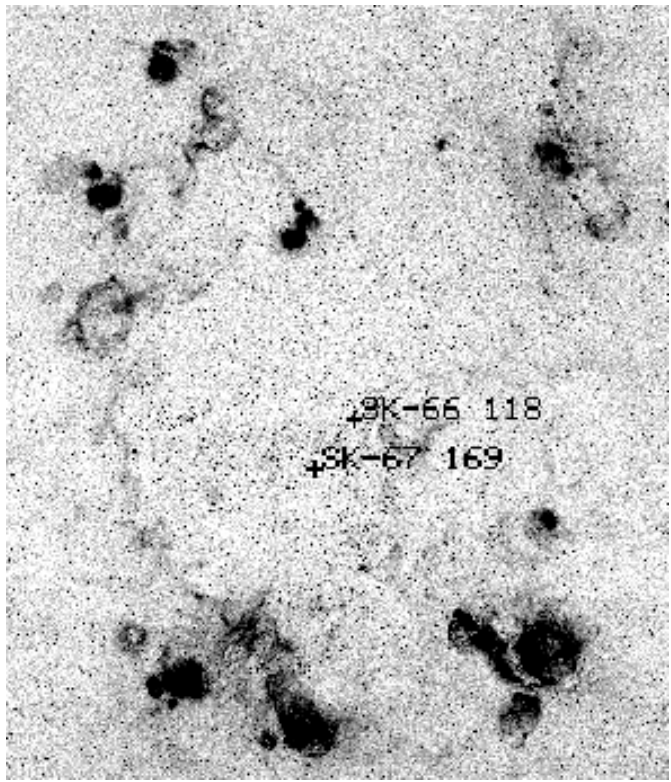}}
\caption{H$\alpha$ image of supergiant shell \lmcv\ made from a scan of a 
photographic plate taken with the Curtis Schmidt 
telescope at Cerro Tololo (Kennicutt \& Hodge 1986). The location of the 
HST target stars are marked with plus signs.}
\end{figure}

Initially we relied on the IUE (Boggess et al. 1978) 
to acquire ultraviolet spectra, 
with the first observations scheduled in 1986.
We found the resulting data extremely useful for various reasons 
(Section 8 and Bomans et al. 1990, Domg\"orgen et al. 1993), 
but due to the limited collecting area and the restricted signal to noise 
ratio achievable with IUE, we could not convince ourselves 
that the data were adequate to yield reliable column densities for a 
hot gas component inside \lmcv. 
We therefore decided to observe a subset of the IUE targets, for limited 
range of wavelengths, with the Goddard High Resolution  Spectrograph (GHRS) 
on the Hubble Space Telescope (HST). 
The data were obtained on 1 July 1993 and 26 December 1993, 
i.e. before the installation of COSTAR.

While the analysis of our HST data was ongoing, 
we detected significant amounts of X-ray emission from 
the inside of \lmcv\ with ROSAT (Bomans et al. 1994). 
Previous X-ray satellites had produced data in which diffuse soft X-ray 
emission seemed to be discernible from the general area of \lmcv, 
but that radiation was too weak and the measurements were 
too noisy to establish proof (see discussion by Bomans et al. 1994). 
Also, Chu et al. (1994) detected significant amounts of \civ\ absorption 
associated with the supergiant shell LMC\,3.
Their data are taken from an IUE survey for high velocity \civ\ absorption, 
and the relatively easy detection of \civ\ in LMC\,3 contrasts with 
our difficulty of firmly establishing this component in \lmcv\ using the same 
instrumentation. While some of the stars in the Chu et al. (1994) survey 
are quite hot, and might have produced some 
circumstellar \civ, the possibility that the physical conditions within the two 
largest LMC supergiant shells are different clearly arises. 
Any models for the supershells would then have to explain 
the variation in these absorptions between otherwise quite similar systems. 

In the present paper we discuss the results from our spectroscopic 
observations toward two stars inside \lmcv\ with the HST, and from five stars 
inside \lmcv\ as observed with the IUE.
The location of the stars observed with HST is indicated in a sketch of 
the region (Fig. 1), for the others we refer to Dom\"orgen et al.\,(1995). 
The actual data and reduction procedures will be described in section 2, 
while in section 3 we briefly discuss 
the stellar features in the HST spectra as far as they are relevant to the analysis 
of the interstellar absorption features. 
In section 4 we turn to the low ionized gas, while in section 5 we 
derive information about \civ\ from the spectra. Section 6 describes 
the spectral restoration algorithm adopted, as well as the results derived from the 
spectra processed in this manner. Section 7 introduces an analysis technique which involves 
profile fitting. Results thereof are then compared with those of sections 5 and 6, and provide 
a useful cross check. A confrontation between the HST results and those from IUE is 
provided in section 8. Finally, section 9 is devoted to a discussion of the consequences of 
our observations.

\section{Observations and data reduction}

We used the Goddard High Resolution Spectrograph (GHRS) aboard the Hubble 
Space Telescope (HST) to take spectra of 
\ssn\ and \sea\ in two spectral ranges each. We selected these two 
stars, because the IUE high dispersion spectra prove that the stars are 
located inside \lmcv\, and their projected location in the very center 
of \lmcv\, should provide the largest path-length through the 
supergiant shell.

The technical data of those four HST observations are summarized in 
Table 2.
The two wavelength ranges chosen contain a large number of interstellar 
absorption lines representing gas of quite different ionization properties. 
In the spectral region from 1530 to 1570 \AA\ lines of \civ\ and 
of C\,I, C\,I*, and C\,I** are expected, 
while in the spectral range from 1285 to 1325 \AA\ 
absorption lines of O\,I, Si\,II, O\,I*, Ni\,II are located.
C\,I represents the neutral gas where the lines from the excited levels 
sample the densest parts of that gas. 
Si\,II and O\,I sample normal neutral gas and, due to the relatively large 
intrinsic optical depth of the transitions, also the more dilute portions. 
Ni\,II traces the densest parts of the neutral gas. 
\civ\ in absorption represents the presence of highly ionized 
and hot gas. 

The aberated PSF of the (pre-COSTAR) HST, in combination with the large 
science aperture (LSA) of the GHRS, 
results in a spectral point spread function consisting of a narrow 
($\sim$ 16 \kms) core and a 60 \kms\ wide halo. 
This, of course, is a considerable degradation relative to the nominal 
resolution of the G160M grating which should be 
about 11.5 \kms\ (Duncan 1992). 
On the other hand, using the small science aperture would have involved
a considerable loss of light, which we considered unacceptable in view of the 
faintness of our targets. As will be shown in the later sections,
the choice of the LSA provides us with sufficient 
signal to noise to detect and analyze weak -- hot gas -- absorption lines.

Other observational parameters are the use of the default substep pattern 4, 
as well as the standard FP-split. The so-called comb-addition 
was used to reduce the effects of the fixed pattern noise of the 
Digicon diode array.
On-board Doppler compensation, which corrects for the orbital motion of the 
spacecraft, was working correctly during all our integrations.
The wavelength calibration was improved using spyballs as described in the 
HST data handbook. 

The spectra were resampled to a linear wavelength scale with 0.0355 \AA\ 
pixel$^{-1}$ for the spectral region of the Si\,II line and 
0.0345 \AA\ pixel$^{-1}$ for the spectral region of the \civ\ doublet.
Figure 2 shows the spectra; the S/N at the continuum level in each spectrum 
is given in Table 2. 

The spectra were normalized by fitting Legendre polynomials to the continuum. 
The steeply rising part of the strong stellar P Cygni profiles had to be 
fitted independently from the other parts of the spectra. In these 
subregions Legendre polynomials, also fitted with the IRAF `continuum' 
again gave a tight fit to the stellar continuum.
Fig. 3 shows the profiles of the interstellar lines after normalization.

\begin{table}
\caption[]{Data on the GHRS spectra}
\begin{tabular}[]{lcrcr}
\hline
Star & $\lambda_{c}$ & $t_{\rm int}$[s] & S/N & $v_{\rm rad}$(LSR)\\
\hline
\ssn & 1305 \AA & 1305.6 & 24 & +258\\
\ssn & 1550 \AA & 6528.0 & 38 \\
\sea & 1305 \AA & 435.2  & 14 & +275\\
\sea & 1550 \AA & 2176.0 & 24 \\
\hline
\end{tabular}
\end{table}

\begin {table}
\caption[]{Data on the IUE spectra}
\begin{tabular}{lrll}
\hline
Star & SWP & $t_{\rm exp}$[m] & Observer\\
\hline
\ses & 27896 & 361 & de Boer \& Koornneef\\
\sea & 28100 & 380 & de Boer \& Koornneef\\
\sns & 27888 & 308 & de Boer \& Koornneef\\
\ssn & 43299 & 380 & Bomans\\
     & 44362 & 406 & Bomans\\
\snn & 31339 & 850$^a$ & Garmany\\
\sss &  6967 & 240$^a$ & Conti\\
     &  8011 & 255$^a$ & Stickland\\
\hline
\end{tabular}

$^a$ taken from IUE archive
\end{table}

The IUE spectra were obtained in full shift exposures from VILSPA, 
three spectra were taken from the IUE archive. 
The relevant data are given in Table 3. 
The spectra cover the wavelength range of about 1150 \AA\ to 2000 \AA. 
The basic reduction of the spectra was done with standard routines at the 
IUE observatory and the spectra were further processed in Bonn 
using MIDAS.
This included rebinning, slight filtering of the oversampled data to reduce 
the noise, and transformation to a velocity scale. 
Part of the IUE data has been presented in a study of the dynamics of the 
neutral gas in \lmcv\ (Domg\"orgen et al. 1995).
In 1986, a spectrum of \ses\ at about 5$\arcmin$ from \sea\ was also obtained; 
this spectrum turned out to be much weaker then we expected, and the 
achieved S/N ratio does not justify an analysis of the absorption features.

\begin{figure*}
\def\epsfsize#1#2{0.9\hsize}
\centerline{\epsffile{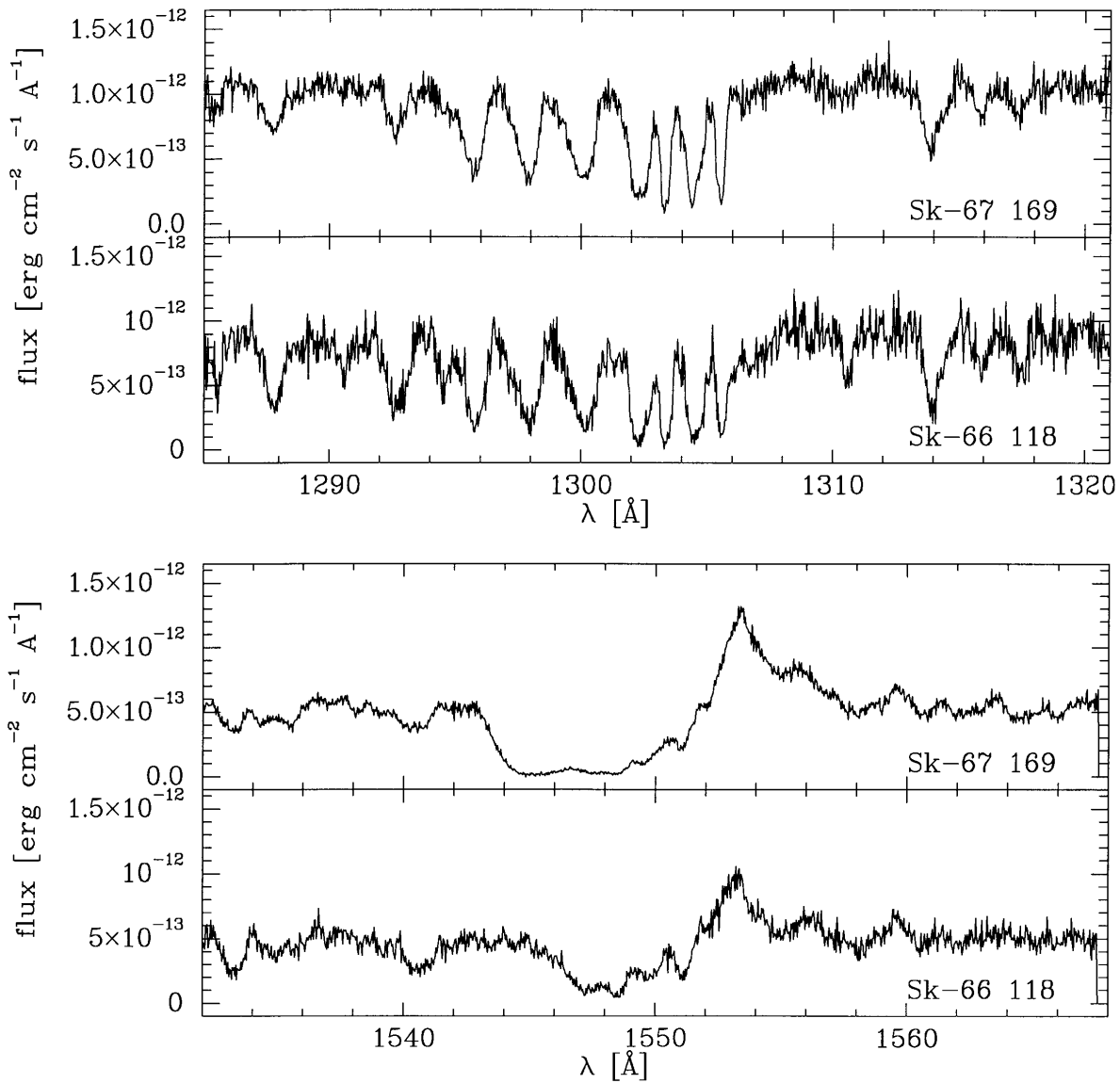}}
\caption{Spectra obtained with the HST for \ssn\ and \sea\ are shown 
for the wavelength ranges 1285 to 1321 \AA\ and 1532 to 1568 \AA. 
Here the strong stellar Si\,III lines (1294.543, 1296.726, 1298.891+1298.960, 
1301.146, 1303.320 \AA) and the \civ\ 
P Cygni like line (centered at 1549 \AA) can be recognized. 
The sharper structures are due to interstellar absorption lines of 
O\,I, Si\,II, and Ni\,II in the former wavelength range and due to the \civ\ 
doublet in the latter wavelength range. 
Note that the interstellar absorptions are double due to galactic gas 
near 0 \kms\ and to LMC gas near 270 \kms\ (LSR).}
\end{figure*}

\section{Stellar lines}

In addition to the interstellar absorption features that are the primary focus of our present investigation, both the spectral regions 
we observed with the HST-GHRS contain strong stellar lines. 
The spectral region around 1550 \AA\ is dominated by the P Cygni 
profile of the \civ\ doublet, while in the spectral region around 1300 \AA\ 
strong absorption lines of particularly Si\,III (1294.543, 1296.726, 
1298.891+1298.960, 1301.146, 1303.320 \AA; see e.g. Wollaert et al. 1988) 
are present, as visible in Fig. 2. 
These stellar lines in our HST spectra can be used for a spectral 
classification independent of the one given in Domg\"orgen et al.\, (1995), 
based on high dispersion IUE spectra of relatively low S/N, 
and the optical classification from the literature.  
The strength of the stellar lines observed in our HST spectra appear to 
agree with  the classification listed in Domg\"orgen et al. (1995), 
confirming a higher temperature for \ssn\ than for \sea. 

Using the 5 strong Si\,III lines around 1300 \AA\ we 
determined the radial velocity of our target stars. 
The results are 258$\pm$5 \kms\ for \ssn\ and 275$\pm$7 \kms\ for \sea. 
The uncertainty given here is the internal uncertainty of the measurement.
The actual error may be slightly higher due to the blending of the 
galactic component of the interstellar O\,I and Si\,II with the two 
highest wavelength Si\,III lines and residual effects of the wavelength 
calibration of the spectra using the GHRS spy-balls.

The derived radial velocities are typical for 
LMC stars in the field of supergiant shell \lmcv. The radial velocity 
of \sea\ coincides with the velocity of a broad H\,I emission component
at this position (Domg\"orgen et al. 1995). 
The radial velocity of \ssn\ is 20 \kms\ lower than the H\,I 
components listed in Domg\"orgen et al. (1995), but coincides with 
interstellar absorption components 
found in the IUE spectra of this star.

\begin{figure*}
\def\epsfsize#1#2{0.8\hsize}
\centerline{\epsffile{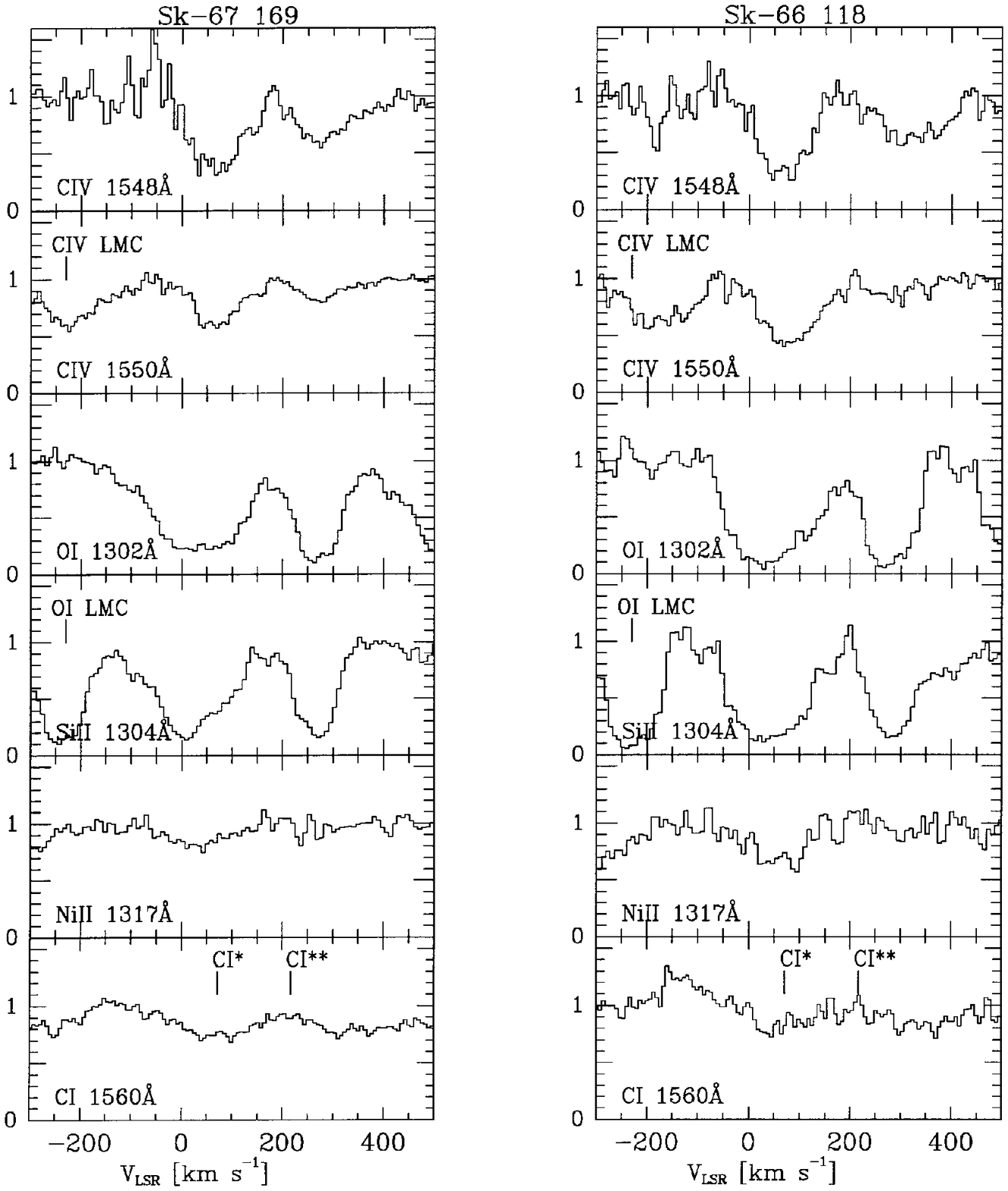}}
\caption{The interstellar absorption lines present in the HST spectra 
have been normalized to the stellar continuum and are plotted against 
velocity (LSR), with from top to bottom 
O\,I 1302.168, Si\,II 1304.370, Ni\,II 1317.217, 
and the two lines of the \civ\ doublet at 1548.195 and 1550.770 \AA. 
Note the absorption due galactic gas between $-$50 and +180 \kms\ and 
by LMC gas between +200 and +350 \kms. 
The strong noise at low velocity in the C\,IV 1548 profiles results 
from the very low stellar background flux in the absorption component of the stellar P Cygni 
profile.}
\end{figure*}

\section{The low ionized gas}

Interstellar absorption by O\,I and Si\,II is clearly detected in the 
spectra of both stars over a velocity interval from 30 to 150 \kms, 
while the wings to negative velocities ($-$30 \kms) are probably the 
effect of a blend with stellar Si\,III lines.
The interstellar absorption components in the above velocity range belong most 
probably to the galactic halo (Savage \& de Boer 1981, de Boer et al. 1990), 
and are known since the first IUE 
observations of LMC stars (Koornneef et al. 1979). 
Additional strong absorptions in O\,I and Si\,II are seen 
at velocities between 200 and 330 \kms\ and are associated with 
interstellar gas belonging to the LMC (see e.g. de Boer \& Savage 1980 and 
Domg\"orgen et al. 1995).

The weak interstellar line of Ni\,II at 1317 \AA\ is not convincingly 
detected in our spectra. While the spectrum of \snn\ shows two 
weak dips at about the right velocity for the two dominant LMC 
components visible in O\,I and Si\,II, absorption at galactic velocities 
only consists of a broad, weak depression in the continuum without clear 
indication of narrow interstellar absorption components. From the 
IUE data we know that the column density of galactic gas is at least 
comparable to the LMC column densities. 
In view of the known lower metal 
abundance of the diffuse gas in the LMC (e.g. de Boer et al. 1987) 
the two weak spectral structures at LMC velocity 
are unlikely to be Ni\,II detections. 
In the spectrum of \sea\ no Ni\,II absorption is visible at LMC velocities.
At velocities compatible with galactic disk and halo gas, two weak dips 
are visible on top of a gentle depression of the continuum. 
Again, comparison with the absorption strength of the Si\,II and O\,I lines 
casts doubt on the reality of the lines. 
The halo components are generally located on 
the linear part of the curve of growth, therefore 
their strength is linearly correlated with the column density. 
The Ni\,II lines at halo velocities (if they are real) are 
much stronger than at galactic disk velocities. 
This would imply Ni\,II overabundance in the halo clouds along this line 
of sight, a highly improbable situation.
More probable is a stellar Ni\,II line of this relatively `cool' B2 star.

C\,I is present in our spectra over the full velocity range 
(0 to 300 \kms) in each line, but the limited 
spectral resolution of (pre-COSTAR) GHRS does not allow us to disentangle 
C\,I and the 4 excited C\,I lines in this spectral region.

The two lines of excited states of O\,I at 1304.86 and 1306.03 \AA\ are not 
detected in our spectra.

\section{Interstellar \civ}

The stellar \civ\ P-Cygni profiles  prominently 
visible the two lower panels of Fig. 2 show significant spectral structure 
which is due to interstellar absorption indicating narrow absorption lines 
superimposed on the smooth stellar continuum. 
After fitting the steeply rising continuum (see Sect.\,2), 
four narrow interstellar absorption components are nicely visible (Fig. 3). 
Their central wavelengths correspond to two components of the \civ\ 
doublet each with the galactic and LMC gas velocity structure. 
For \ssn\ the central absorption velocities are 57 and 270 \kms, while 
for \sea\ the features are centered at 63 and 280 \kms. 
The strong absorption near 60 \kms\ is known to be due to hot gas in the halo 
of the Milky Way (Savage \& de Boer 1981). 
We associate the absorptions near the systemic velocity of the 
LMC with hot gas inside the supergiant shell \lmcv.

We will make the analysis of the \civ\ lines in 3 steps. 
First, we will investigate  the equivalent widths of the 
absorbing components (see below).
An analysis of the actual profiles in terms of velocity dependent 
optical depth follows (Sect. 5.2). 
Finally, we apply a deconvolution-reconstruction algorithm to the spectra,
taking into account the actual PSF of the GHRS. 
The optical depth analysis of Section 5.2 is then repeated for these 
`sharpened' spectra, and the results compared. 
Since this procedure requires a fairly elaborate discussion, 
it is presented in a separate section (Sect. 6). 

\subsection{\civ\ equivalent widths and column densities}

Under the simplifying assumption that the observed equivalent widths 
are due to a single Galactic plus a single LMC component, we can 
determine their \civ\ column densities using the curve of growth approach. 

The LMC absorptions have equivalent widths of about 80 m\AA\ and 200 m\AA\ 
for the two lines of the doublet, which is close to the theoretical 
-- optically thin -- ratio of two for these lines.
In the optically thin approximation, the equivalent widths translate readily 
into \civ\ absorption column densities  
of $N({\rm \civ})$ = 5~10$^{13}$ cm$^{-2}$ for both lines of sight.

In contrast, the two galactic halo absorption components 
are about equally strong (300 m\AA) and are therefore saturated. 
Thus, only a lower limit for the column density of these high ions 
can be found, with $N({\rm \civ})\,> 10^{14}$ cm$^{-2}$. 
Note, however, that this result suffers some additional uncertainty due to 
the location of the galactic component of the 
\civ\ line at 1548 \AA\ near the bottom of the stellar \civ\ absorption line.  

The applicability of the curve of growth method is severely limited if 
an apparently optically thin observed profile actually consists of 
narrow and unresolved 
components, which individually could well be significantly saturated. 
For this case, the `apparent optical depth' approach offers some relief, 
and we now discuss the results of this method. 

\subsection{\civ\ column densities from apparent optical depth}

If an observed absorption profile shows spectral structure, we can apply 
the velocity resolved optical depth method (e.g. Sembach \& Savage 1992).
In this method, the data points of the observed line profiles are converted 
to apparent optical depth: 

\begin{equation}
{\tau}_a(v) = ln \left(\frac{I_{cont}(v)}{I_{a}(v)}\right),
\end{equation}

\noindent
and thereafter to apparent column density:

\begin{equation}
N_a = \left(\frac{m_e c}{\pi e^2}\right) \frac{1}{\lambda f}~{\tau}_a 
~~{\rm [atoms~cm^{-2}~(km\ s^{-1})^{-1}] }.
\end{equation}

\noindent
Fig. 4 shows the apparent optical depth profiles for the spectral region
around the \civ\ lines.

The condition that the narrowest intrinsic structure of the spectral 
line is fully resolved is, at a GHRS resolution of about 11 \kms\, not 
necessarily fulfilled for lines due the low ions. 
As is well known (see e.g. $\zeta$ Oph, de Boer \& Morton 1974), 
such absorptions may be due to clouds with 
b-values as small as 1 \kms.

However, for the absorptions of \civ\ the direct transformation 
from absorption depth into optical depth is valid since the \civ\ ion 
exists normally only in gas of temperatures above 10$^5$\,K. 
The thermal width then exceeds about 20 \kms. 
Only if the \civ\ ions were to exist in gas which has cooled down to 10$^4$ K 
without having recombined (in the case of rapid adiabatic expansion, 
which `freezes' the ionizational structure; e.g. Breit\-schwerdt \& 
Schmutzler 1994) could the intrinsic structures be narrower 
(see next section). 

For \ssn\ the profiles of the two lines agree with each other to 
within the errors, implying no hidden saturated components. 
Decomposing the profile in components, the \civ\ column density can therefore 
be accurately determined to $1~10^{14}$ cm$^{-2}$ for the 70 \kms, 
$1.5~10^{13}$ cm$^{-2}$ for the weak 140 \kms\ component, and 
$6~10^{13}$ cm$^{-2}$ for the 270 \kms\ component.

For \sea\ the situation is different. The optical depth profile 
clearly indicates saturation over most of the velocity range, 
but in particular for the component of the galactic halo at about 70 \kms\, 
limiting the column density determination there to a lower limit of 
$1~10^{14}$ cm$^{-2}$. 
For the LMC component the somewhat lower S/N in this spectrum 
becomes a factor, but it is nevertheless clear that in some parts 
of the profile unresolved, saturated components are present. 
Therefore the \civ\ column density for the LMC part of the line-of-sight 
derived from the \sea\ spectrum of $6~10^{13}$ cm$^{-2}$ is somewhat uncertain 
and probably has to be interpreted as lower limit. 

\begin{figure}
\def\epsfsize#1#2{1.0\hsize}
\centerline{\epsffile{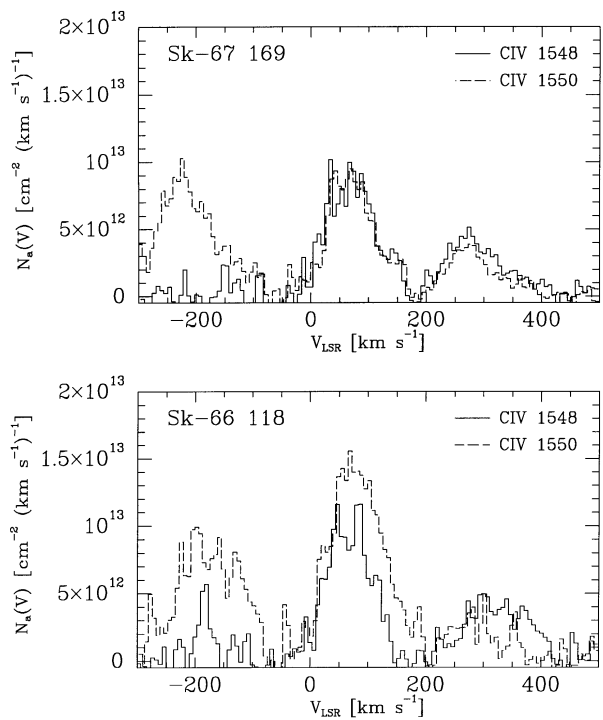}}
\caption{The absorption seen in the \civ\ lines of Fig. 3 is transformed 
into apparent optical depth profiles and after appropriate conversion 
the column density profile is obtained as shown. 
In those locations where the column density profile, $N(v)$, 
of the 1548 \AA\ line deviates from that of the 1550 \AA\ line, 
saturation is present in the stronger of the two lines. }
\end{figure}

\section{Spectral Restoration}

While the interstellar absorption lines of interest are clearly visible 
in our spectra,  their velocity structure (as indicated from our IUE 
high dispersion spectra of these stars) is compromised by the asymmetric 
line spread function of the pre-COSTAR GHRS. 
As the signature imposed by the GHRS on the observed spectra is well 
known and temporally stable, the intrinsic spectral line profiles can be 
recovered to some extent. Not only are we interested in achieving a match 
to the intrinsically narrowest lines, but we would also like to know 
whether the broad wing to higher velocities seen in the line profile of 
the component at LMC velocity is real or due to the 
line spread function (LSF) of the telescope-spectrograph combination. 

To recover some of the spectral resolution, and to make optimal use of the 
high S/N in our spectra, we used a modified Lucy algorithm with entropy 
constraint (Walsh \& Lucy 1993) to deconvolve our spectra. The method 
conserves the flux, while the entropy constraint helps to keep the 
noise from amplifying. However, finding a balance between an optimal 
sharpening and generating unphysical, spurious, spectral structure remained 
the major challenge. 

We made several runs with different parameters. 
The results we liked best were obtained 
with 30 iterations and a very small entropy constraint.   
We used for this process exclusively the data of \ssn, because 
the spectra have a much higher signal to noise ratio. Initial tests with 
the \sea\ spectra showed that for these datasets the resolution gain is 
not outweighed by the much higher noise level. Scientifically, 
this a not a big loss, 
because we determined already in Section 5.2 that the interstellar C\,IV
absorption lines in \sea\ are saturated, and they will thus not show 
additional structure at higher spectral resolution. 

Figure 5 shows the results of the restoration process in comparison 
to the original data of \ssn. Normalization was done for both data 
sets in the same way as described in section 2.

\begin{figure*}
\def\epsfsize#1#2{0.8\hsize}
\centerline{\epsffile{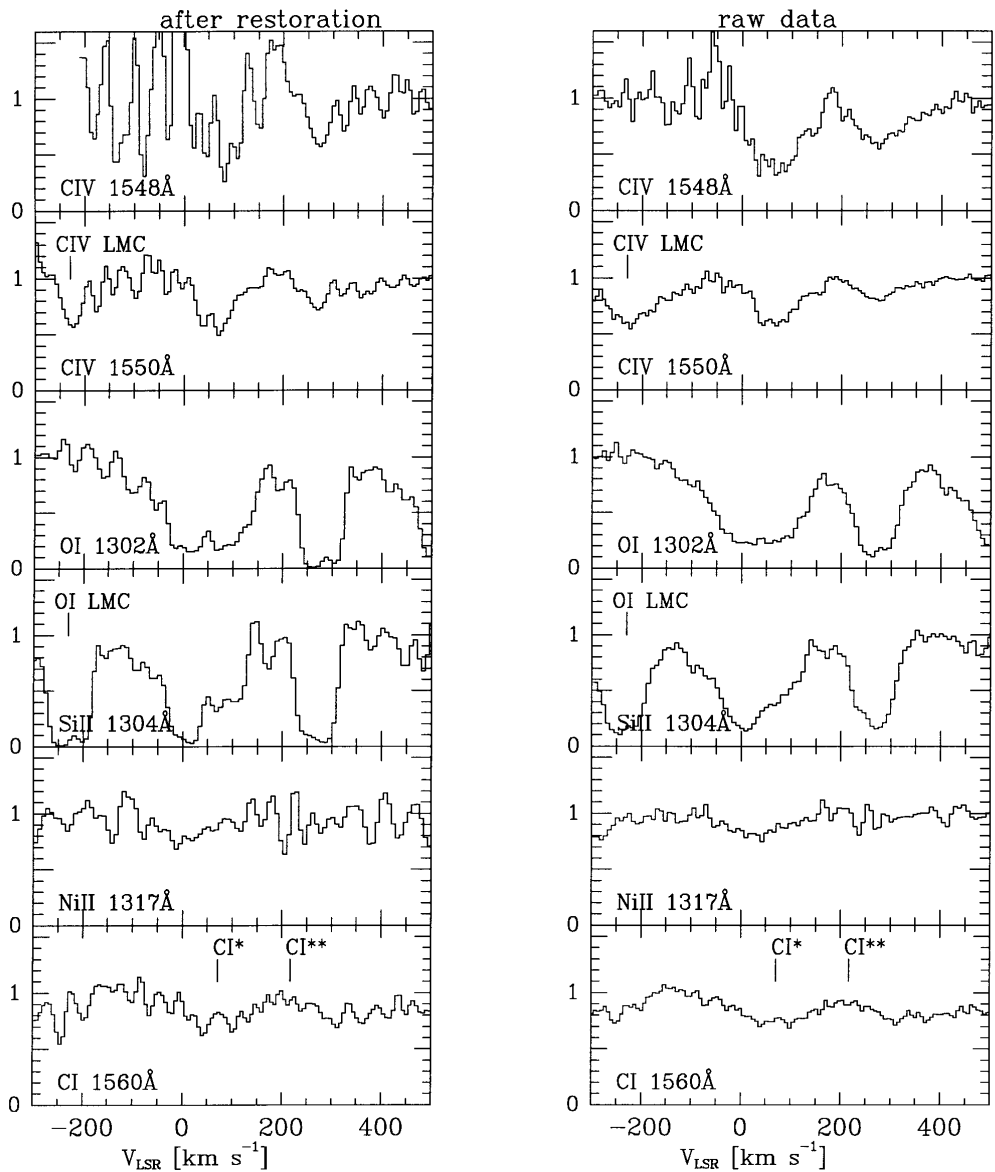}}
\caption{Interstellar line profiles in the \ssn\ spectra {\bf{after}} and 
before application of the resolution restoration routine.
}
\end{figure*}

The interstellar absorption lines are much better defined in the 
deconvolved spectra, clearly showing a gain in spectral resolution, 
especially by correcting for the broad wings of the LSF. The 
line profile of Si\,II provides a good example. 
The actual spectral resolution is difficult to determine, because in 
our spectra no isolated, narrow interstellar line is present; 
using the only slightly blended lines it can be estimated to be 
about 16 \kms. This value is very close  to the theoretical limit given 
by the width of the LSF core (Duncan 1992).

\subsection{Low ionization stage lines}

The velocity structure of the neutral gas as traced by Si\,II 
can now be interpreted as corresponding to interstellar clouds 
at 20, 65, 110, 250 and 290 \kms. 
In the IUE spectra these components were also found (see e.g. 
Domg\"orgen et al. 1995), except that the two components 
at 275 and 300 \kms claimed there merge into a single component 
near 290 \kms\ in the HST spectra. 
The difference is likely due to the higher S/N in the HST spectra. 

The velocity pattern from the O\,I line agrees reasonably well with that 
from Si\,II for the galactic components, 
having radial velocities of 20, 70, and 100 \kms. At LMC 
velocities a slightly different pattern is visible with absorption 
components at 260 and 305 \kms. 
This may indicate that the 300 \kms\ component in the IUE spectra has 
a higher column density in O\,I than in Si\,II. 
One possible explanation for this difference could lie in a variation 
of elemental abundances due to different depletion into dust grains, 
which could occur if these velocity components correspond to 
spatially distinct interstellar environments (i.e., inside or outside 
supergiant shell \lmcv\ respectively).
Clearly  much more data would be needed to test this speculation.

The broad depression of the continuum of the Ni\,II line at 1317 \AA\ 
is now resolved into the principal velocity components at 
galactic velocity. At LMC velocity still no Ni\,II is convincingly detected. 

\subsection{The \civ\ absorption in \ssn\ }

The largest gain from the deconvolution process is visible in the 
\civ\ lines. The high-velocity wing is transformed into a weak additional 
absorption component and the main absorption dip is now a nice, 
symmetric, single component profile.
This shows clearly that the high velocity wing was an artifact of 
the line spread function of the pre-COSTAR HST-GHRS instrument. 

Following the procedure given earlier, we converted the deconvolved 
spectrum into apparent optical depth, and then calculated the apparent 
column density profiles.
These profiles are given in Fig.6.
The velocity components of the \civ\ line can now be traced 
as 35, 73, 145, 277, and 318 \kms.
We must note that the relatively poor signal to noise in the \civ\ 
1548 profile is the result of the location of this line near the bottom 
of the stellar \civ\ P-Cygni profile of \ssn. 

\begin{figure}
\def\epsfsize#1#2{1.0\hsize}
\centerline{\epsffile{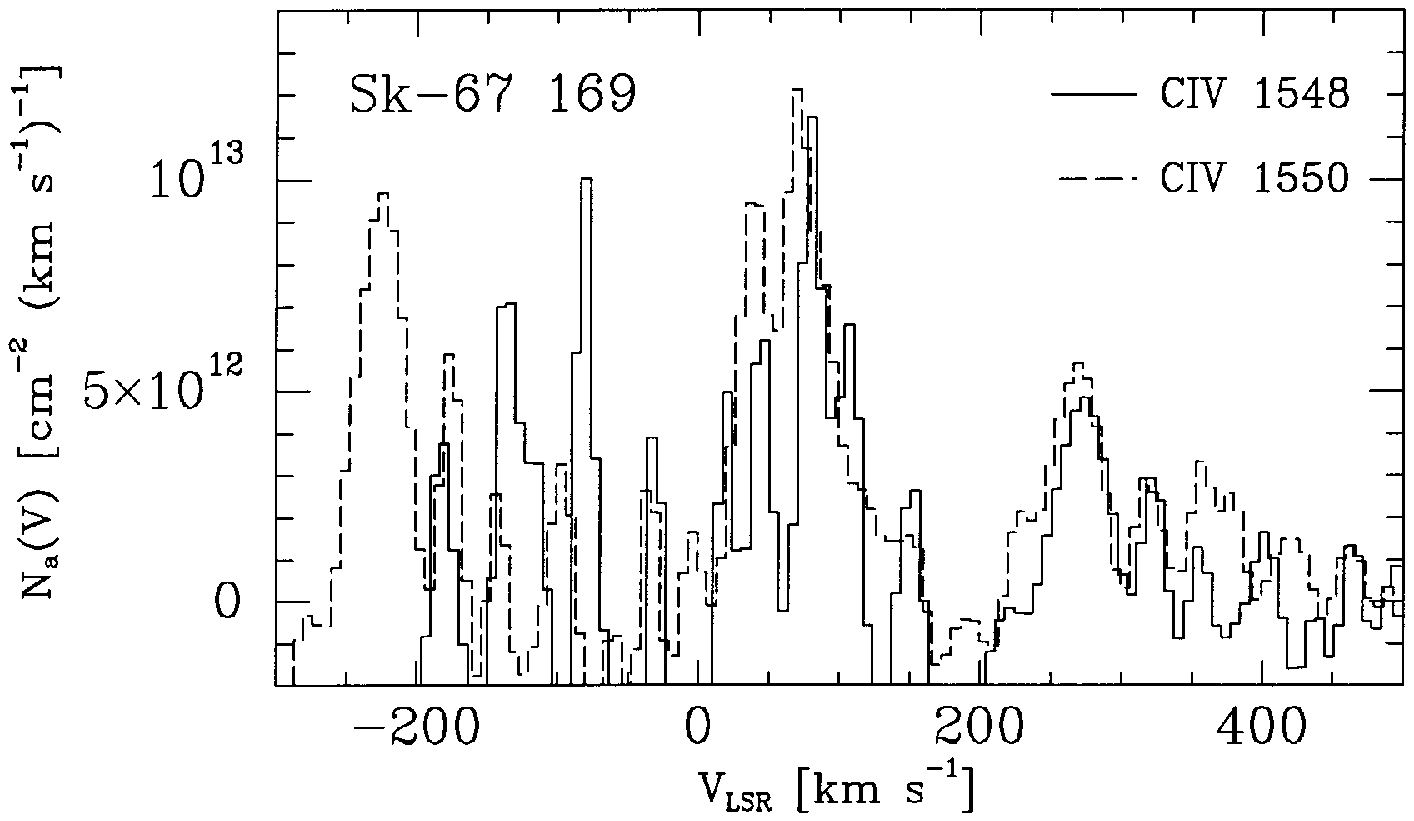}}
\caption{Apparent column density profiles of the \civ\ of \ssn\ AFTER 
application of the resolution restoration software.}
\end{figure}

The galactic 35 \kms\ is somewhat uncertain due to low S/N, but if 
real, then  the line is  saturated. For all other components the 
profiles of the two \civ\ lines agree well, therefore accurate 
column densities can be derived by integrating the apparent column 
density profile in the velocity intervals of the components.
Table 4 shows the results, which agree well with the results of 
the integration of the raw spectrum and the curve-of-growth results, 
proving that the spectral restoration did not significantly affect 
the flux scale. 
When comparing the column densities one has to keep in mind that due 
to the spectral restoration more components could be measured individually 
as result of the resolution of the broad profiles in the 
unprocessed spectra.

The galactic \civ\ component at 73 \kms\ agrees  well in velocity and column 
density with the results of Savage \& de Boer (1981) toward other LMC stars.
The \civ\ column densities and velocities of this absorption component on the 
two lines of sight (which are 17$\arcmin$ apart)  
indicate that the origin of this component is located in the halo of our 
Galaxy and not in the LMC. At the distance of the LMC the absorbers would 
have to be one cloud with a diameter larger than 250 pc to produce 
these very similar absorptions on our two lines of sight. 
A cloud in the halo of our galaxy is an order of magnitude closer in 
space and the apparent homogeneity is then expected. 
A weak component at 110 \kms\ is not well resolved in the restorated spectrum 
of \ssn, but seems to be present in both lines of the \civ\ doublet. We 
therefore do not give a individual column density of the component.
The same comments about the physical location as for the 73 \kms\ component 
apply.

The 145 \kms\ \civ\ component on the line of sight to \ssn\ has no counterpart 
in the spectrum of \sea. One possibility is that this is an additional small 
high-velocity cloud (HVC) only intercepted 
on one line of sight. Similar velocities are known in the general area 
(see de Boer et al. 1990). 
However, contrary to the two well known HVCs at about 60 and 130 \kms\ 
this \civ\ absorption component has no counterpart in Si\,II and O\,I. 

The two C\,IV components at LMC velocity are well resolved now, which 
enable us to use the actual profiles of the absorptions to derive 
additional information. The stronger absorption at 277 \kms\ 
is symmetric and has a full width at half maximum of 40 \kms\ 
(39 \kms\ for \civ\ 1548 and 41 \kms\ for \civ\ 1550). This is 
much larger than the spectral resolution defined by the core of the 
LSF (11.5 \kms; Duncan et al. 1992). Even if we did not recover 
the full intrinsic spectral resolution 
of the GHRS G160M grating, our estimated  
resolution of 16 \kms\ implies that this 277 \kms\ component is 
easily resolved. 

We can now use the line width to derive a kinetic temperature of 
the absorbing \civ\ gas, if we assume pure thermal 
broadening. The full width half maximum, FWHM, is given by 

\begin{equation}
FWHM = 2 \sqrt{ln~2}~\sqrt{\left(\frac{2~k~T}{m}\right)} 
\end{equation}

\noindent
and therefore

\begin{equation}
FWHM = 0.215~ \sqrt{\frac{T}{A}}~~~{\rm [km\ s^{-1}]} 
\end{equation}

\noindent
with temperature $T$, the mass of the atom $m$, 
the Boltzmann constant $k$, and the atomic mass number $A$ (Savage 1987). 
For 10$^5$ K, the temperature of the peak ionic abundance under 
thermodynamic equilibrium (e.g. Shull \& van Steenberg 1992), 
the \civ\ absorption lines would have a width of 20 \kms.  
To this we must add the effect of the smearing due to the 
instrumental profile (16 \kms, see above), 
near \civ\ bringing the total width to 27 \kms. 
The observed \civ\ component at 277 \kms, 
if interpreted as due to such thermal gas, 
would imply a plasma of 1.4 $10^6$ K. 
At this temperature the ionic abundance of \civ\ would be low. 
The observed width therefore must mean that the gas is 
either in non-equilibrium conditions or a that the line width has a 
non-thermal origin, such as blending of components or velocity gradients. 
A detailed discussion of the \civ\ results will be given in Sect. 9.

\begin{table}
\caption[]{C\,IV column densities}
\begin{tabular}{lrrr}
\hline
Star & $v_{\rm LSR,C\,IV}$ & $N_{\rm \civ,r}$ & $N_{\rm \civ,d}$  \\
\hline
\ssn &  35 &             & $> 2~10^{13}$\\
     &  73 & \raisebox{1.5ex}[-1.5ex]{$1~10^{14}$} & $ 6~10^{13}$\\
     & 145 & $1~10^{13}$ & $ 7~10^{12}$\\
     & 277 &  & $ 3~10^{13}$\\
     & 318 & \raisebox{1.5ex}[-1.5ex]{$6~10^{13}$}  & $ 6~10^{12}$\\
\sea &  73 & $> 1~10^{14}$ & \\
     & 130 & $> 3~10^{13}$ & \\
     & 280 & $ 6~10^{13}$ & \\
\hline
\end{tabular}

Notes: r, from raw spectra; d, from deconvolved spectra
\end{table}

\section{Profile fitting}

An alternative way to analyze the profiles of interstellar lines observed
with GHRS before COSTAR is described in Lu et al. (1994). They fitted their 
spectra with Voigt profiles which were smeared with the aberated PSF.
We used the MIDAS context CLOUD for the same process and used the 
the line spread function given by Duncan (1992). Because the spectra 
of \ssn\ have clearly better signal to noise ratio and the lines are 
not all saturated, we used only this object. 

The fit supports the results from the deconvolution: 
after plotting a best fit one-component model over the spectrum the weak 
additional \civ\ component at 145 \kms\ clearly shows up. 
The high velocity wing of the \civ\ line turns out to be 
partly due to the instrumental profile and partly due to a real feature. 
We tried to make a fit with more than the three components for the 
galactic halo \civ\ and the two components for the LMC absorption. 
This did not work well, which may be an effect of slight deviations of the 
actual LSF of our observation from the standard  LSF given by Duncan (1992). 
These differences can easily be induced by the target not being perfectly 
centered in the aperture. 
The results of the spectral restoration do not unambiguously allow 
more components into the fit.

\section{The IUE spectra and comparison with HST}

The spectra obtained with the IUE are of only very moderate signal-to-noise 
and have a spectral resolution equivalent to slightly more than 20 \kms. 
A detailed analysis of the velocity structure of the low 
ionized lines is given in Domg\"orgen et al. (1995). 

While there are indications of interstellar \civ\ absorption 
in IUE high dispersion spectra to targets inside \lmcv\ (Bomans et al. 1990, 
Domg\"orgen et al. 1993), the interpretation is severely limited 
by the low S/N of the available IUE high dispersion spectra. 
\ses\ and also \sea\ have been dropped for the analysis, 
because the spectra turned out to be too noisy.
Table 4 shows the results. 
The most convincing detection is in the spectrum of \sss, with 
$N({\rm \civ})$ = 5.0~10$^{13}$ cm$^{-2}$ at a velocity of 270 \kms.
\sss\ is unfortunately an O6 star, and a significant amount of 
the detected \civ\ may be due to ionization of 
the circumstellar surroundings by the star itself. 
This circumstellar contribution needs more detailed analysis before 
any conclusions can be derived from this interesting line of sight. 
The same is true for the line of sight to \snn.
HST observations of such hot stars and detailed modeling of the circumstellar 
contribution of the star will be presented elsewhere.
The other IUE spectra give upper limits for the \civ\ column density 
in agreement with our detections presented in this paper.

Due to the large spectral range covered by a SWP spectrum of the IUE, 
additional information can be derived from the Si\,IV and N\,V doublets 
at 1400 and 1240 \AA. Results are collected in Table 5. 

Si\,IV is possibly detected but only in the weaker line of the doublet. 
This is similar to the situation of C\,IV in the HST spectra, 
where the stronger line of the P\,Cygni doublet leaves too little signal for 
a clear detection of the superimposed interstellar absorptions. 
However, the velocity of the dip is 260 \kms, about 20 \kms\ lower 
than the \civ\ absorption in the HST spectrum. Additionally, 260 \kms\ 
is exactly the stellar radial velocity of \ssn, which implies 
that the Si\,IV absorption line probably originates in the circumstellar 
environment of this B1 star. 
The association of the Si\,IV absorption with that of \civ\ is therefore 
very questionable  and no temperature for the interstellar hot gas 
on the \ssn\ line of sight can be derived from this ratio.

Assuming equilibrium electron collisional ionization and solar abundances 
we can use the calculations of Shull \& van Steenberg (1982) 
together with the observed column densities 
to estimate the gas temperature.
For \ssn, using the \civ\ column densities from the HST spectrum 
and the limits for Si\,IV and N\,V derived from the IUE spectrum, 
we get $N$(\civ)/$N$(Si\,IV) $\sim$ 0.5 
and $N$(N\,V)/$N$(\civ) $\le$ 5.

The limit for the N~V to \civ\ ratio is very high, consistent with 
a temperature of less or equal to $1~10^6$ K,  
in agreement with a thermal interpretation of most of the \civ\ line width.

\begin{table}
\caption[]{Column densities of high ions detected in IUE spectra at 
$v_{\rm rad}$ $\sim$ 260 \kms}
\begin{tabular}[]{lrrrr}
\hline
Star & $N$(Si\,IV)& $N$(\civ) & $N$(N\,V) \\
\hline
\ssn & $ 4.0~10^{13}$ & $\le 2.0~10^{14}$ & $\le 1.5~10^{14}$  \\
\sns & $\le 1.5~10^{13}$ & $\le 6.0~10^{13}$ & $\le 1.0~10^{14}$  \\
\sss &                   & $5.0~10^{13}$     &               \\
\snn &                   & blemish           &               \\
\hline
\end{tabular}
\end{table}

\section{Discussion}

We have found substantial column densities of \civ\ in the direction of stars 
inside the supergiant shell \lmcv. The \civ\ absorption is seen at 
277 \kms\ for the stronger component. 
This velocity is marginally smaller than that of the bulk material 
(stars and neutral gas) of \lmcv\ having $v_{\rm LSR}$ = 287 \kms\ 
(see Domg\"orgen et al 1995). 

  From the lack of \civ\ gas in the velocity range between 250 and 200 \kms\ 
on both lines of sight we can 
conclude that a blow-out of \lmcv\ into the LMC halo has, at present, 
probably not taken place at the front side of \lmcv. 

The observations of absorption lines from gas inside the superbubble 
have a topographical bias; since we do not know at what depth the stars lie 
inside \lmcv\ we more likely sample gas of the near side of the superbubble. 
The detected velocities indicate therefore that the \civ\ gas 
is coming toward us. If we assume that the gas comes from a mixing 
layer (e.g. Slavin et al. 1993) between X-ray emitting gas inside \lmcv\ 
and the cool shell, then the superbubble is still expanding 
with a very small velocity of about 10 \kms, well fitting to the 
results of Domg\"orgen et al. (1995) for the front side of \lmcv.  
In this case the large width of the line would rather be explained by 
turbulent motions in the conducting interface than by thermal broadening. 

The other possibility for the origin of the 280 \kms\ \civ\ absorption 
is the interior of \lmcv\ itself. Interpreting the line-width as 
purely thermal the resulting temperature of $1.4~10^6$ K agrees within the 
errors very well with the temperature of the X-ray emitting plasma inside 
\lmcv\ having a temperature of $2.4~10^6$ K (Bomans et al. 1994). 

Depending on the location of the stars inside \lmcv, 
we can derive an average gas density of \civ.
With a column density $N$(\civ) = $6~10^{13}$ cm$^{-2}$, 
the average density of \civ\ for a 1 kpc column (star at the rear side of 
\lmcv) or for a 500 pc column (star at the center of \lmcv) is 
$n$(\civ) = 2 or $4~10^{-9}$ cm$^{-3}$, respectively. 
At the temperature of the 
peak ionic abundance of \civ\ 
and the metallicity of the LMC this corresponds to 
$n_{\rm H}\ = 1~10^{-4}$ cm$^{-3}$. 
For the temperature of $2~10^6$ and LMC metallicity this corresponds to 
less than $1~10^{-8}$ cm$^{-3}$  in H. 
The average density of the hot gas derived 
from ROSAT data of $n_{\rm e}$ = $8.0~10^{-3}$ cm$^{-3}$ 
effectively excludes 
the high temperature implied by a thermal interpretation of the line width 
because the observed \civ\ column density is larger by orders of magnitudes.
We can therefore conclude that the line width of the \civ\ line is at least 
partly due to bulk motions of \civ\ clouds.


The other result of the crude estimation is that, 
even at the temperature of the peak ionic abundance of \civ, 
the filling factor of \civ\ is smaller than that of H. 
This again supports an origin of the the \civ\ absorption from clouds 
inside \lmcv\ and from a conducting layer on the front side of 
the expanding supergiant shell. 
The filling factor will be even smaller, if we assume that the hot gas 
inside \lmcv\ is not at LMC abundance of about 1/3 solar, but more 
metal enriched due to the supernova explosions inside the supergiant shell.

Using the density estimate for the conducting front hypothesis, it 
becomes clear that one layer of pc size would imply a density of the order
of $n$ = 1 cm$^{-3}$, which would lead to rapid cooling. Many layers 
along the line of sight seem to be a more realistic description.

Interestingly the column density of \civ\ is the same for both lines of sight, 
which (if the stars are at equal depth in \lmcv) may point to a common origin. 
This may be one large cloud or a large collection of randomly distributed 
small clouds.

In the course of the discussion above we 
found indications for a population of \civ\ clouds inside \lmcv. 
This may be a population of cool clouds which 
survived the passage of several shock-waves inside \lmcv, or even was created 
by the passage of these shock waves through a clumpy ISM 
(Ferrara \& Shchekinov 1993). 
These clouds are pushed around randomly and will all have interfaces with 
the hot plasma, in which most of the \civ\ is produced. Some 
contribution of quiescently cooling gas (e.g. Slavin \& Cox 1992) to the 
\civ\ column density will be present as well.
In this scenario our lines of sight will intersect several of this \civ\ 
layers. 
The average velocity of the created line will be near to the rest velocity 
of the gas in the region of \lmcv, or maybe at somewhat lower velocity due 
to the LMC~4 depth sampling bias explained above.  
The line-width of the resulting \civ\ line would be dominated by the bulk 
motion of the clouds and not by thermal broadening. 

If these clouds (i.e. the mixing layers) are abundant inside \lmcv, as 
implied by our data, 
this has strong implications for the evolution of supergiant shells, 
leading to a stronger cooling inside, which limits their age and therefore 
their size. 
The need for an additional energy source becomes much more severe.
More lines of sight have to be sampled to investigate the size and 
the filling factor of these \civ\ producing clouds to ensure that 
our two examples are not exceptional. 

The second \civ\ component visible in both lines of sight, either as wing 
to high velocities or as second component (apparent after deconvolution) 
has to have a different origin because of its large velocity difference to 
the rest velocity at this location of the LMC.
Its velocity is also higher than any velocity component of low ionized 
or neutral gas seen in absorption in our spectra. 
This excludes the possibility that this gas is related to the rear side of 
\lmcv, which appears to blow out into the halo of the LMC 
(Domg\"orgen et al. 1995). 
The high velocity gas appears to have similar 
velocity structure and column density in both observed lines of sight, 
as derived from the similarity of the high velocity wings in the not 
restorated spectra. 
This implies that the gas `cloud' has an minimal size of 250 pc leading to 
large energy requirements independent of the actual 3-d structure. 
One possible explanation is that we detected a blast wave from a 
supernova which exploded inside the cavity of \lmcv. Such an 
event should lead to a faint, fast moving filament of enhanced emission 
(McCray 1988), describing the observation well. 
Interestingly, faint filamentary structure is also visible in the ROSAT 
images of \lmcv, which may have the same creation mechanism. 
Alternately the absorption could also be caused by a supernova fragment 
as proposed to exist in supergiant shells by Franco et al. (1993).

Based on early IUE spectra de Boer \& Savage (1980) claimed that the LMC 
is surrounded by a hot halo, just like the Milky Way (Savage \& de Boer 1981). 
The argument rested on the strengths of the \civ\ line and its velocity 
(see review of de Boer 1984). The velocity of the halo gas was near 220 \kms, 
a value much smaller that what we have detected toward \lmcv.
In the HST \civ\ profiles one can see absorption extending down to about 
that velocity but no specific absorption component is present. 
In a recent paper, containing a detailed analysis of all archived and 
a significant number of new IUE high dispersion spectra of LMC stars, 
Wakker et al. (1996) show that there is no global corona of the LMC, 
but that a patchy one is still consistent with the data. 
Unfortunately from the present HST data we cannot derive further constraints 
about a hot corona of the LMC.

\acknowledgements
We thank Jeremy Walsh and Adeline Caulet at the ST-ECF for help during 
preparation of the phase two of the observations and with the basic steps 
of the HST data reduction. 
DJB likes to thank the ST-ECF staff for their hospitality during two stays 
at Garching. 
We thank especially Jeremy Walsh for letting us use his 
spectral restoration routines and for discussions about the handling of 
LSA data suffering from the aberated PSF.
We thank You-Hua Chu and Mordecai Mac Low for comments and discussions. 

DJB and EKG have been financially supported partly through the 
DFG Graduiertenkolleg `Magellanic Clouds'.  DJB thanks the Alexander von 
Humbold Foundation for support as part of their Feodor Lynen Fellowship 
program.

\enddocument